\documentstyle[12pt,aasms4]{article}

\begin{document}

\title{Does Every Quasar Harbor A Blazar?}

\author{Feng Ma and Beverley J. Wills}
\affil{McDonald Observatory, University of Texas,  Austin, 
TX 78712; feng@astro.as.utexas.edu, bev@astro.as.utexas.edu} 

\begin{abstract}

Assuming there is a blazar type continuum in every radio-loud
quasar, we find that the free-free heating due to the beamed 
infrared continuum can greatly enhance collisionally excited lines,  
and thus explain the stronger CIV $\lambda$1549 line emission observed 
in radio loud quasars. We further predict that the CIV line should
show variability {\it not} associated 
with observed continuum or Ly$\alpha$ variability. 

\end{abstract}


\clearpage

\section{Introduction}

The dichotomy between radio-loud and radio-quiet QSOs
remains one of the greatest
puzzles in studies of AGN. While their optical-UV spectra appear
similar, 
quasars (we use the term ``quasar'' for radio loud QSO) 
have CIV $\lambda$1549 line emission $\sim$50\% stronger
compared with radio-quiet QSOs (Francis, Hooper, \& Impey 1993). 
Marziani et al. (1996) find that the luminosities of CIV and H$\beta$ are
strongly correlated in radio-quiet QSOs, but only weakly 
correlated in quasars. Some observed line profile differences could 
also be the result of an additional 
beamed optical-UV synchrotron continuum in quasars, 
but not in radio-quiet QSOs (Wills et al. 1993; Francis et al. 1993). 

Unified schemes for quasars hold that
there is a beamed synchrotron continuum in
every quasar (e.g., Urry \& Padovani 1995). 
If the beam is toward the observer, the quasar is 
seen as a blazar. Blazars are known to be highly variable on time 
scales from days to years. For example, BL Lacertae
brightened by 3 magnitudes in R band during the 
outburst in the summer of 1997 (Bloom et al. 1997). 
The outburst lasted for a few months, 
and during the outburst it shows intraday and even intrahour variation 
of $\sim$1 magnitude. 

Bregman et al. (1986) found that the Ly$\alpha$ line flux varied in
proportion to the continuum flux density in the blazar 3C446. 
Stephens \& Miller (1984) found that the CIII] $\lambda$1909 line flux varied
with the continuum in 3C446 and suggested that
the beamed continuum might affect the emission line spectra. 
Koratkar et al. (1998) found no variation in the flux of Ly$\alpha$ emission line 
in HST spectra of the blazar 3C 279 while the 
continuum varied by a factor
of 50. Cohen et al. (1987) found possible Mg II line variability in 
the blazar AO 0235$+$164. 
Corbett et al. (1996) found the H$\alpha$ line in BL Lac to be 
variable, but not as a result of the beamed continuum. 
The emission line variability in blazars is not
easy to measure because of the large contribution from the 
variable continuum.  
Most previous studies of AGN spectral variability
have focused on objects known to have large continuum variability 
and concentrated on the emission line variations following the 
{\it observed} continuum variations (e.g., Maoz et al. 1994). 

In this Letter we investigate the interactions between  the 
beamed  synchrotron radiation and the BELR. Although the beam may illuminate 
only a small portion of the Broad Emission Line Region (BELR), 
it can provide a large flux of infrared 
photons that heat the gas by the 
free-free mechanism (Ferland et al. 1992), greatly enhancing the CIV 
$\lambda$1549 line emission. This enhanced emission can significantly 
contribute to the
line flux from the entire BELR, explaining the stronger line emission 
seen in quasars. 
The CIV line emission 
is thus expected to vary in response to the rapid and 
large amplitude variations of the 
blazar continuum, although we may not see the continuum variation if 
it is beamed away from us.  

\section{Additional Beamed Continua in Quasars}

Theoretically, a simple blazar synchrotron continuum is 
beamed with a half opening angle 
${\theta}={\sin}^{-1}(1/\gamma)$   
for a Lorentz factor $\gamma$. 
The fraction of the sky area illuminated by the beam is $1/2{\gamma}^2$, 
or ${\sim}2\%$ for ${\gamma}=5$, and ${\sim}5\%$ for ${\gamma}=3$.
The real beaming angle could be larger (Lind \& Blandford 1985), 
and can be estimated from the observed fraction of blazars among quasars
as a whole (e.g., Urry \& Padovani 1995).

If the beamed synchrotron continuum originates closer to 
the central black hole than 
the BELR (${\sim}0.1$ pc in luminous quasars, 
see e.g., Netzer 1990) then it may affect
the emission line spectrum. In NGC\,1275 the VLBI jet 
probably originates within 0.1 pc at $43$ GHz (Krichbaum et al. 1992).   
The higher frequency synchrotron 
radiation of interest (${\sim}10^{12}$ Hz) is likely to 
originate closer to the black hole.  
Some theoretical models predict infrared synchrotron emission 
originating ${\sim}0.1$ pc from the black hole  
and that lower frequency emission originates farther away 
from the center (see e.g., Marscher 1996). 

The beamed synchrotron continuum provides a high flux 
of IR photons because of
its steep spectrum, which extends to mm wavelengths.  
In contrast, the normal QSO
continuum provides only weak IR flux, with much of 
the observed IR emission
almost certainly arising from dust beyond the 
BELR (Barvainis, Antonucci, \& 
Coleman 1992).  Moreover, if the observed IR continuum 
in radio-quiet QSOs were
within BELR distances it would significantly overpredict 
the observed CIV strength.

\section{Models}

Ferland et al. (1992) showed that a high flux of IR photons 
incident on BELR clouds could enhance the flux of the CIV and OVI 
emission lines by free-free
heating of the BELR.
As discussed above, a strong IR central continuum is 
unlikely in radio-quiet
QSOs.  However, for BELR gas exposed exposed to the 
beamed synchrotron radiation, 
free-free heating is the most important 
heating process.

The heating rate is proportional to (e.g., Netzer 1990)
\begin{equation}
 \displaystyle 
  F_{\nu}e^{-\tau_{\nu}(ff)}T^{-1/2}_eN_eN_i{\nu}^{-3}(1-e^{-h\nu/kT_e}).
\end{equation}
Collisional excitation results in an important emission line cooling 
process in optically thick BELR gas, and
the rate is proportional to 
\begin{equation}
 \displaystyle
  N_e\Omega_{ij}e^{-E_{ij}/kT_e}T^{-1/2}_e, 
\end{equation}
where $\Omega_{ij}$ is the collision strength. For CIV $\lambda$1549, 
$E_{ij}\sim 8$eV, and the cooling rate increases with temperature. 
At ${\sim}16$ eV, this line intensity stops 
increasing with $T_e$ because the collision cross sections become smaller 
for higher velocity electrons. 
Roughly speaking, the IR heating leads to a higher electron temperature, 
which results in stronger CIV $\lambda$1549. 

We use a fixed $L_{\rm agn} = 10^{46}$ erg s$^{-1}$  for an  
isotropic AGN continuum and use the spectral energy distribution (SED) 
of Mathews \& Ferland (1987) except that the submillimeter break 
is at 10\micron.  The beamed continuum is conservatively assumed to have 
 $f_{\nu}~{\propto}~{\nu}^{-1}$ from 
$10^{-3.5}$ to $10^{-2}$ Ryd, and 
$f_{\nu}~{\propto}~{\nu}^{-2}$ from $10^{-2}$ to $10^6$ Ryd. 
At lower energies, $f_{\nu}~{\propto}~{\nu}^{2.5}$. 
We assume that  the beamed continuum has constant
intensity within the small opening angle $2\theta$, and 
use $L_{\rm beam}$ to represent the apparent luminosity as seen by
an observer within the beam. 
Adding $L_{\rm beam}$ essentially
does not change the ionization parameter because of the steepness of 
the spectrum. 

Solar abundances are assumed for the BELR gas. The calculations 
are done using CLOUDY (Ferland 1996). 
We first try a simple BELR model ``A'' resembling a model given by 
Ferland \& Persson (1989). 
They use a fixed hydrogen density $n_H=10^{9.5}$ cm$^{-3}$, 
and a fixed ionization
parameter $U = 10^{-0.5}$. Because we are using 
$L_{\rm agn}=10^{46}$ erg s$^{-1}$ 
for the input continuum,  $U = 10^{-0.5}$ constrains the cloud distance to
$r=10^{17.67}$cm. The column density for the cloud is set to be
$10^{25.5}$ cm$^{-2}$. 

In the more realistic model ``B'' we follow the model of Goad, O'Brien, \& 
Gondhalekar (1993),  
in which discrete clouds are distributed from $10^{17}$ cm to $10^{19}$ cm 
from the continuum source. The density starts from $10^{13}$ cm$^{-3}$, 
and has a 
power law distribution $n~{\propto}~r^{-2}$. The column density is 
set to be $N(r)~{\propto}~r^{-4/3}$, starting from $10^{23}$ cm$^{-2}$ at 
the inner radius. The differential covering factor varies as 
${\propto}r^{-1/6}$, 
to fulfill the mass conservation 
of the clouds. The calculated emission line luminosities are integrated 
from inner to outer radius. 

\section{Results}

In Tables 1 and 2 we list the calculated emission line luminosities for
models A and B. 
For the H$\beta$ line we give the logarithm of luminosity in units of 
erg s$^{-1}$, and the other line
luminosities are given as the ratio to H$\beta$. 

Column 2 is the result for  $L_{\rm agn}=10^{46}$ erg s$^{-1}$ with no 
$L_{\rm beam}$; 
column 3 is for $L_{\rm agn}=10^{46}$ erg s$^{-1}$ mixed with 
$L_{\rm beam}=10^{46}$ erg s$^{-1}$, and  $L_{\rm beam}$ 
affects only 2\% of the sky area. A small covering factor (${\sim}0.1$) for the
clouds is assumed everywhere in the sky, including in the beam. 
Columns 4 and 5 correspond to 
$L_{\rm beam}=10^{46.5}$ and $10^{47}$ erg s$^{-1}$. 
The observed line ratios in column 6 are taken from Netzer (1990). 

From Tables 1 and 2 it can be seen  that among strong lines,  the 
CIV $\lambda$1549 is enhanced by ${\sim}$50\% when
$L_{\rm beam}=10^{47}$ erg s$^{-1}$,  
while hydrogen lines show little change. 
The CIII $\lambda$977 is also strongly dependent on the 
IR heating. These two lines actually set up an upper limit 
of $L_{\rm beam}~{\sim}~10^{47}$ 
erg s$^{-1}$, otherwise they would be too strong. However, we note
that CIII $\lambda$977 is difficult to observe because of 
intervening absorption, especially at higher redshifts. 
Model A predicts a slight enhancement of OVI $\lambda$1035 and
NV $\lambda$1240. We do not consider their enhancement in model 
B, because the strengths of these two lines are not well predicted. 

In Fig. 1 we plot for model A the abundance ratio C$^{3+}$/C, 
and the electron temperature versus the depth into the cloud. 
As expected, the C$^{3+}$ density is not strongly affected by the beamed
continuum, and CIV $\lambda$1549 is an effective coolant. 

In Fig. 2, we give cloud surface temperature and 
various line luminosities versus radius in model B. 
The free-free heating is more important at small radii, since it is 
proportional to the photon flux as shown in eq. (1), while photoelectric
heating by ionizing photons is determined by the hydrogen number
density and the recombination coefficient and does not depend 
explicitly on photon flux (Ferland et al. 1992). 
 
Some line profiles with and without a beamed continuum are compared 
in Fig. 3. We show the profiles here primarily for the purpose of showing 
the enhancement of integrated emission lines flux due to the 
beamed continuum, though affecting only a small portion of the 
BELR clouds. The BELR cloud distribution and velocity field 
are probably anisotropic with velocities having a strong
disk component in addition to a random component (e.g., Wills \& Browne
1986). For simplicity, 
here we consider only spherically distributed clouds with a 
Keplerian velocity field, and the velocity direction is assumed to be
random. The maximum cloud velocity is $10^4$ km s$^{-1}$ at $r=10^{17}$ cm.  
The flat tops of the profiles come from the lowest gas velocity,  
$10^3$ km s$^{-1}$ at $r=10^{19}$ cm. 

\section{Discussion}

The IR-optical continuum of the 
synchrotron radiation is steep, hence 
the fraction of hydrogen ionizing photons it can 
offer is small compared with the normal AGN continuum.  
In addition, the beam illuminates only a small fraction of 
the BELR clouds, while the emission lines we see are integrated
over all the clouds. Consequently, the 
effect on  the H$\alpha$ and Ly$\alpha$ lines is negligible.
For collisionally excited lines such as CIV $\lambda$1549, 
the line emissivity can be enhanced by over two orders 
of magnitude in the beam, and the line intensity integrated
over the whole BELR consequently increases by a factor of $\sim$50\%. 

If there is a beamed blazar type continuum in every radio loud
quasar, we should be able to see CIV line variability
{\it not} associated with observed continuum 
variability or Ly$\alpha$ variability. This phenomenon 
could well have escaped
detection in previous observations because high redshift
quasars have not been extensively monitored. Also, this phenomenon
may be seen only when the blazar continuum is in the outburst stage. 
We are planning observations to search for CIV $\lambda$1549 
line variability among $z~{\sim}~2.5$ quasars, 
which are thought to be less variable compared with low luminosity
AGNs. If the CIV line variability at the level $\sim$50\% is observed, 
while the lines for comparison (Ly$\alpha$, CIII] 
$\lambda$1909 or MgII $\lambda$2798) 
do not change significantly, it 
will be  strong support for our models and for quasar unified schemes. 
In addition, the disk-wind
model (e.g., Emmering, Blandford, \& Shlosman 1992) for the BELR 
will be ruled out, since it does not include any
gas in the polar regions.  

We also expect quasars to have statistically stronger CIV $\lambda$1549. 
The strength of CIV $\lambda$1549
depends strongly on the blazar continuum, and should 
not be correlated well with H$\beta$ line strength as shown by
Marziani et al. (1996). 
The intensity variation
may be seen in the line core or line wings, and thus offer 
information about the gas kinematics in the polar regions, e.g., whether
the gas clouds have lower velocity or have large radial velocities
driven by the jet. 

In reality, the line emission depends on the details of BELR 
geometry.  If the line is strongly beamed backward 
(e.g., Ferland et al. 1992), we may not be able to see
the enhanced CIV emission when the angle between the line of sight
and the beam is large. However, 
the excited lines are at least more isotropic than the beamed 
continuum, and we do expect to see the line variability without
seeing the beam. 

If the phenomenon suggested by us is not detected, this 
will still give strong constraints on models. Three possibilities are left. 
First, the IR synchrotron radiation originates
outside the BELR. Second, the observed blazar continuum variability
may be caused by a change of the jet direction rather than a change 
in strength of the beam (e.g., Abraham \& Carrara 1998). 
Third, there is no
broad line emitting gas in the polar regions. The last possibility
would favor the disk-wind model and may rule out the bloated star model 
(e.g., Alexander \& Netzer 1994). 

Finally we note that if the BELR is made of small clouds 
confined by a hot medium (Krolik, McKee, \& Tarter 1981), then the jet 
might destroy the clouds. However, if the BELR is made of 
bloated stars, 
the jet may enhance the mass loss rate of the stars and increase the 
amount of broad line emitting gas in the beam. 

\acknowledgements

We thank Greg Shields for helpful discussions, and Gary Ferland
for advice and making his code CLOUDY available. We also thank the referee, 
Ski Antonucci, for valuable comments. 
This work is supported by NASA LTSA grant NAG5--3431.

\clearpage

\begin{center} 
{\bf Figure Captions} 
\end{center}
\begin{description}

\item[Fig. ~1] \hspace{1em} Temperature and C$^{3+}$/C profiles inside the cloud in model A. 
The solid lines correspond to the case with $L_{\rm agn}=10^{46}$ erg s$^{-1}$, 
and the dashed lines correspond to the case with $L_{\rm agn}=10^{46}$ erg s$^{-1}$ 
mixed with $L_{\rm beam}=10^{47}$ erg s$^{-1}$. 

\item[Fig. ~2] \hspace{1em} Surface temperature of clouds and 
emission line luminosity versus radius in model B. 
The solid lines correspond to the case with $L_{\rm agn}=10^{46}$ erg s$^{-1}$, 
and the dashed lines correspond to the case with $L_{\rm agn}=10^{46}$ erg s$^{-1}$ 
mixed with $L_{\rm beam}=10^{46.0}, 10^{46.5}, 10^{47.0}$ erg s$^{-1}$
respectively. $L_{\rm beam}$ and $L_{\rm agn}$ 
both illuminate 4$\pi$ steradian. 

\item[Fig. ~3] \hspace{1em} Emission line profiles in model B. 
The solid lines correspond to the case 
with $L_{\rm agn}=10^{46}$ erg s$^{-1}$, 
and the dashed line corresponds to the case with $L_{\rm agn}=10^{46}$ erg s$^{-1}$ 
mixed with $L_{\rm beam}=10^{47}$ erg s$^{-1}$. $L_{\rm beam}$ illuminates only
2\% of the clouds. The fluxes of different lines are arbitrary. 

\end{description}

\clearpage

\begin{deluxetable}{rrrrrrr}
\tablecolumns{7}
\tablewidth{0pc}
\tablecaption{Model A}
\tablehead{
\colhead{} & \multicolumn{5}{c}{$L_{\rm beam}$} & \colhead{} \\
\cline{2-5} \\
\colhead{Line} &0& \colhead{$10^{46}$} & 
\colhead{$10^{46.5}$} &\colhead{$10^{47}$} 
&\colhead{Obs.}
}
\startdata
Ly$\alpha$ 1216          &  25.3 & 25.3 &  25.3&  25.5  & 8-15 \\
CIV $\lambda$1549        &  11.4 & 11.9 &  13.1&  16.0  &  5-8  \\
CIII] $\lambda$1909      &   6.0 &  6.1 &   6.5&  7.5  &  2-4  \\
CIII $\lambda$977        &   0.9 &  0.9 &   1.2&  2.6  & $<$1 \\
H$\beta$ 4861            &  43.59& 43.60& 43.61& 43.61 &  1  \\
H$\alpha$ 6563           &   2.3 &  2.3 &   2.3&  2.3   & 4-6 \\
HeI $\lambda$5876        &   0.2 &  0.2 &   0.2&  0.2   &  0.1-0.2 \\
HeII $\lambda$1640       &   1.1 &  1.1 &   1.1&  1.1   &  0.6 \\
HeII $\lambda$4686       &    0.1&   0.1&    0.1&  0.1  &   0.1  \\
NIV] $\lambda$1486       &    0.7&   0.8&    0.8&  1.0  &  0.7  \\
MgII $\lambda$2798       &   1.2 &  1.2 &   1.3 & 1.3   &  3   \\
SiIV,OIV] $\lambda$1400   &   1.6 &  1.7 &   1.9 & 2.7   &  1.3  \\
OIII] $\lambda$1665      &    3.4&   3.5&    3.8&  5.0  &  0.5  \\
NIII] $\lambda$1750      &   0.9 &   0.9&    1.0&  1.4  &   0.4  \\
OVI $\lambda$1035        &  3.0  &  3.2 &   3.5 & 4.2   &  3 \\
NV $\lambda$1240         &  1.0  &  1.1 &   1.1 & 1.3  &  3  \\
\enddata
\end{deluxetable}

\begin{deluxetable}{rrrrrrr}
\tablecolumns{7}
\tablewidth{0pc}
\tablecaption{Model B}
\tablehead{
\colhead{} & \multicolumn{5}{c}{$L_{\rm beam}$} & \colhead{} \\
\cline{2-5} \\
\colhead{Line} &0& \colhead{$10^{46}$} & 
\colhead{$10^{46.5}$} &\colhead{$10^{47}$} 
&\colhead{Obs.}
}
\startdata
Ly$\alpha$ 1216          & 37.4  &37.4  &37.4    & 37.6&  8-15  \\
CIV $\lambda$1549        & 5.9   &6.4   & 7.4    &10.1 &  5-8    \\
CIII] $\lambda$1909      & 2.4   &2.4   & 2.4    &2.5  &  2-4    \\
CIII $\lambda$977        & 1.1   &1.4   & 2.2    &5.1 &  $<$1  \\
H$\beta$ 4861            & 43.99 &44.00 & 44.00  &44.01&  1    \\
H$\alpha$ 6563           & 4.7   &4.7   & 4.7    &4.6  & 4-6    \\
HeI $\lambda$5876        & 0.7   &0.7   & 0.7    &0.7  &0.1-0.2  \\
HeII $\lambda$1640       & 2.4   &2.4   & 2.4    &2.4  &  0.6   \\
HeII $\lambda$4686       & 0.3   &0.3   & 0.3    &0.3  &  0.1   \\
NIV] $\lambda$1486       & 0.1   &0.1   & 0.1    &0.1  &  0.7   \\
MgII $\lambda$2798       & 9.7   &9.8   & 9.9    &10.2  &  3    \\
SiIV,OIV] $\lambda$1400  & 1.4   &1.6   & 1.9    &2.9 &  1.3    \\
OIII] $\lambda$1665      & 1.1   &1.2   & 1.2    &1.5 & 0.5  \\
NIII] $\lambda$1750      & 0.3   &0.3   & 0.3    &0.4  &   0.4    \\
OVI $\lambda$1035        & $10^{-3}$ &0.002&  0.003 & 0.01  &  3 \\
NV $\lambda$1240         & 0.02   &0.03 &  0.05  & 0.1 &  3  \\
\enddata
\end{deluxetable}

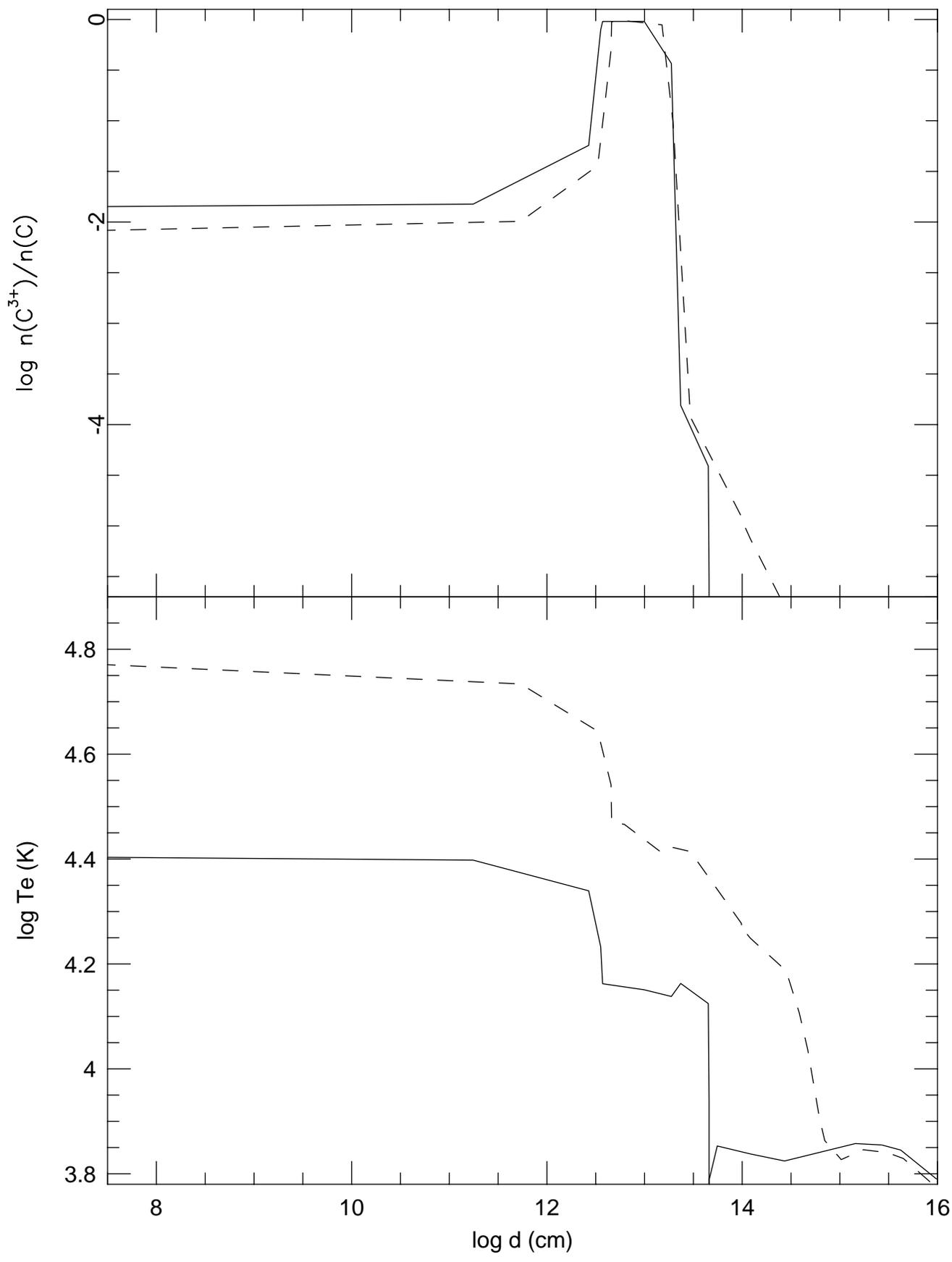
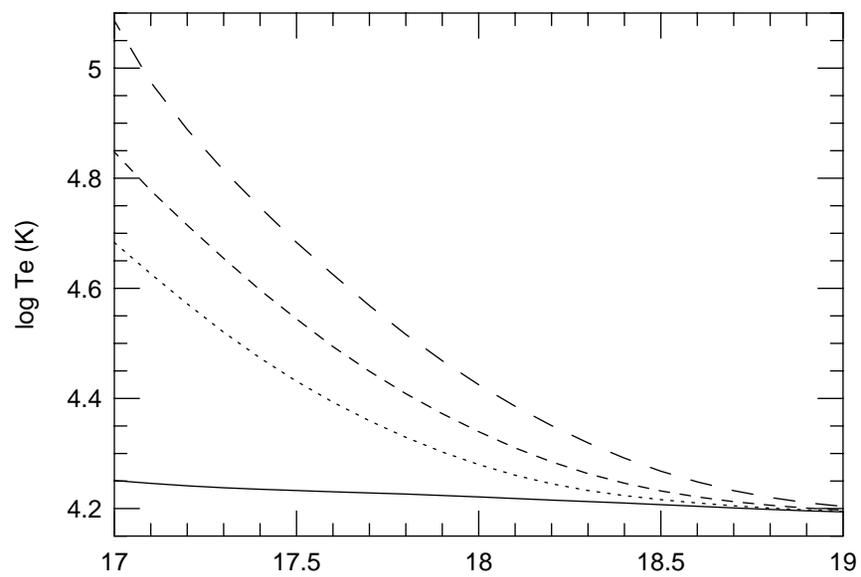
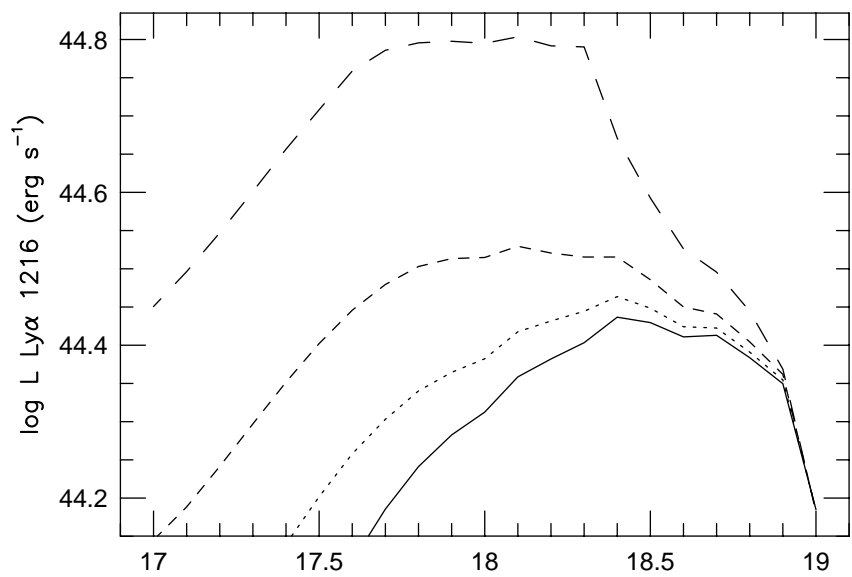
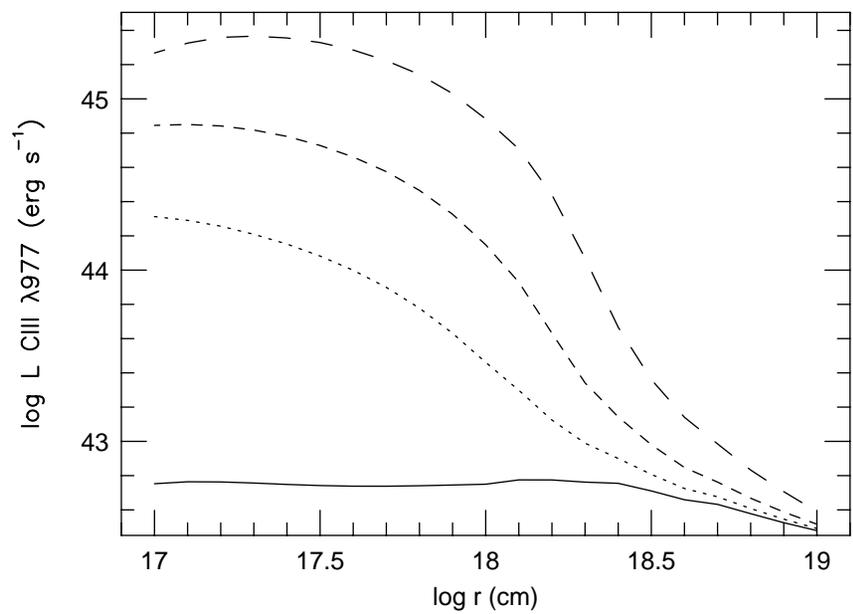
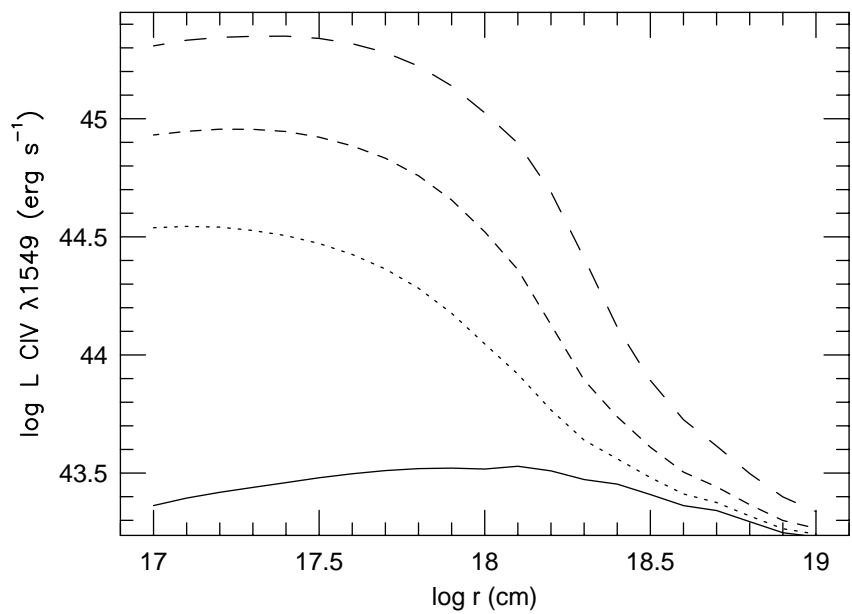
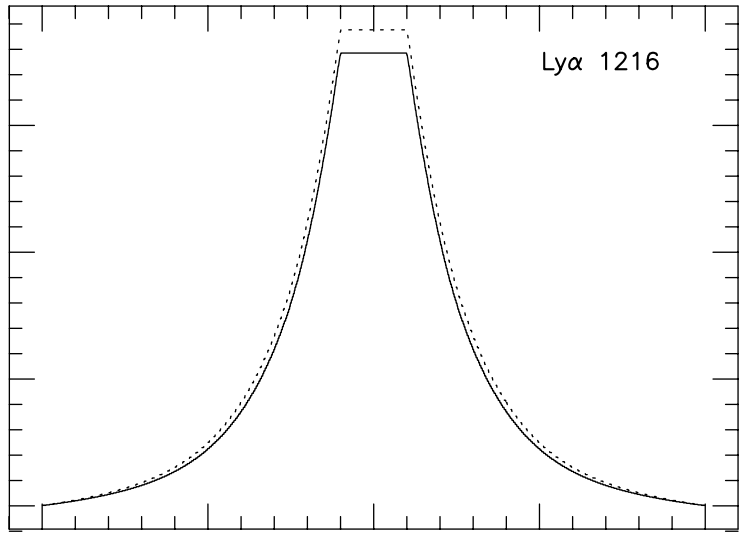
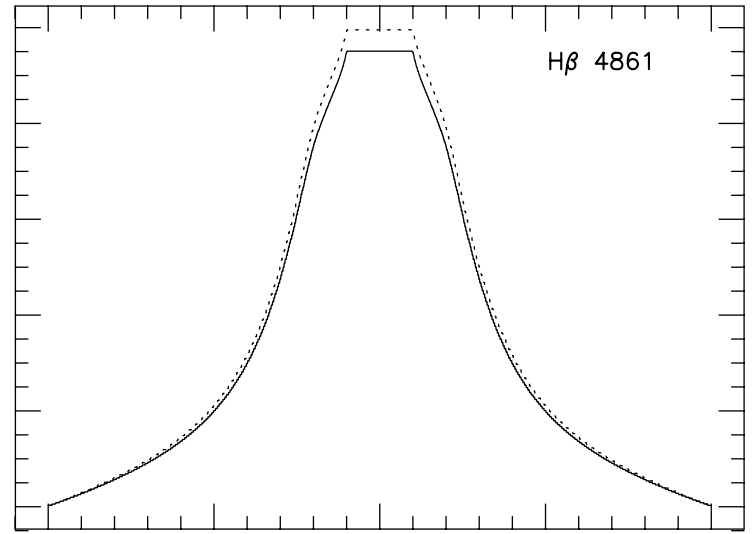
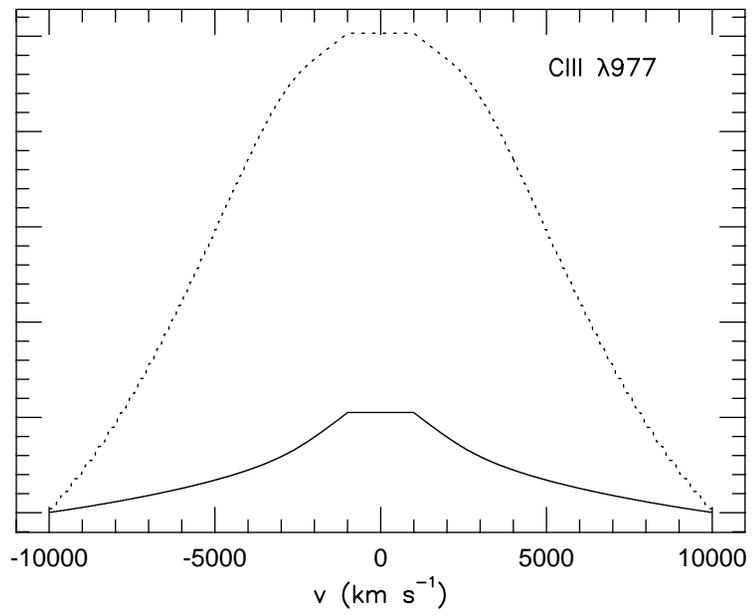
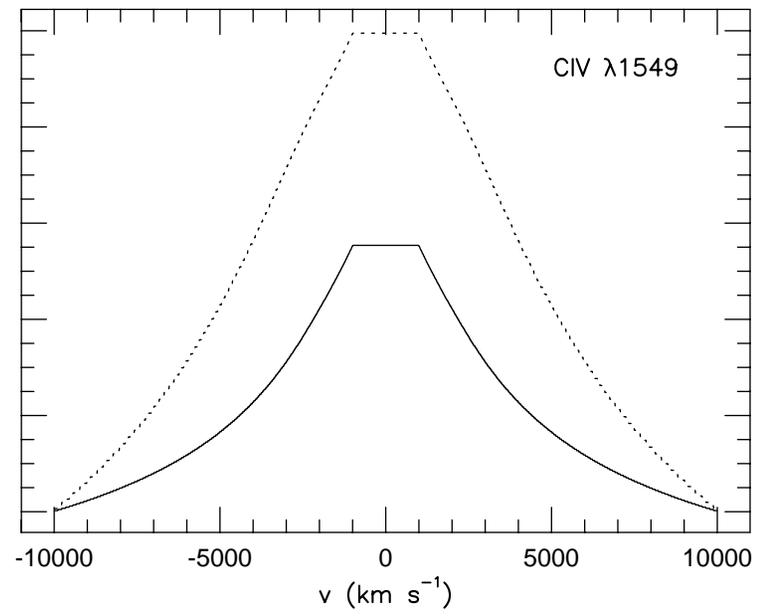
\end{document}